\documentclass[preprint,showpacs,preprintnumbers,amsmath,amssymb,prb]{revtex4-1}
\usepackage{graphicx}
\usepackage{amsmath, amsthm, amssymb}
\usepackage{latexsym}
\usepackage{amssymb}
\usepackage{bm}
\usepackage{dcolumn}
\usepackage{epstopdf}
\usepackage{color}
\usepackage{braket}
\usepackage{mciteplus}

\newcommand{\SHE}{superfluid $^3$He}
\newcommand{\HE}{$^3$He}
\newcommand{\HEF}{$^4$He}

\definecolor{cream}{rgb}{1,1,0.7}
\definecolor{orange}{rgb}{1,0.644,0}

\newcommand{\be}{\begin{equation}}
\newcommand{\ee}{\end{equation}}
\newcommand{\ber}{\begin{eqnarray}}
\newcommand{\eer}{\end{eqnarray}}

\def\ket#1{\mbox{$\displaystyle\vert\,#1\,\rangle$}}

\begin{document}

\title{Superfluid $^3$He in Aerogel}
\date{\today}

\author{W. P. Halperin}

\affiliation{Department of Physics and Astronomy, Northwestern University, Evanston, IL 60660 USA}

\begin{abstract}
Superfluid \HE\ is an unconventional neutral superfluid in a $p$-wave state with three different superfluid phases each identified by a unique set of characteristic  broken symmetries and non-trivial topology.  Despite  natural immunity of \HE\ from defects and impurity of any kind, it has been found that they can be artificially introduced with high porosity silica aerogel.  Furthermore, it has been shown that this modified quantum liquid becomes a superfluid with remarkably sharp thermodynamic  transitions from the normal state and between its various phases. They include new superfluid phases that are stabilized by anisotropy from uniform strain imposed on the silica aerogel framework and they include new phenomena in a new class of anisotropic aerogels consisting of nematically ordered alumina strands.  The study of \SHE\ in the presence of correlated, quenched disorder from aerogel, serves as a model for understanding the effect of impurities on the symmetry and topology of unconventional superconductors.
\end{abstract}

\maketitle

\subsection*{Introduction}

Extensive investigations of  \SHE, since its discovery~\cite{Osh.72,Osh.72b} in 1972, have revealed  unique characteristics of  different thermodynamic phases within a common $p$-wave pairing manifold, including the isotropic fully gapped $B$ phase, the anisotropic $A$ phase, which separately breaks spin and orbital rotation symmetry as well as time reversal symmetry, and the spin polarized superfluid  $A_1$ phase that opens up in the presence of a magnetic field.  The study of the transitions between these phases, and from the normal state, has greatly expanded our understanding of $p$-wave pairing.  In fact,  \SHE\ is the first unconventional superconductor, that is to say one that breaks symmetries of the normal state in addition to gauge symmetry, a class that includes high temperature superconductors and certain heavy fermion materials that are of broad interest in condensed matter physics.  With its discovery was the realization of the pairing theory of Bardeen, Cooper and Schrieffer~\cite{Bar.57} (BCS) for a neutral fermion quantum liquid.  An important distinction between \SHE\ and other unconventional superconductors is that the order parameter of the pure superfluid and its microscopic behavior, including the normal state of fermionic quasiparticles, are very well-established theoretically and experimentally.   Consequently, with \HE\ there is a clear  advantage for the study of the effect of impurities on unconventional pairing.

Investigation of the effects of impurities substituted or inserted into the crystal lattice has been important for understanding disordered conventional BCS superconductors, which in that case arises only from magnetic quasiparticle scattering.~\cite{Gor.60,Abr.61,Gen.66} Unconventional superconductors are much more fragile.  It was shown by Tsuneto~\cite{Tsu.62} that all forms of impurities or material defects can suppress unconventional superconductivity, but within the same basic theoretical framework that had been established by Abrikosov and Gorkov~\cite{Abr.61} for magnetic scattering in conventional superconductors. In fact, observation of the suppression of superconductivity by non-magnetic impurities in new materials is  a useful indication of unconventional pairing as was shown for UPt$_3$~\cite{Dal.95} and Sr$_2$RuO$_4$.~\cite{Mac.98}   Although it is appealing to use this approach with \SHE\ to better understand unconventional pairing, nonetheless it was thought this would be technically impossible since \HE\ is the purest material in nature and does not accept any form of impurity at low temperatures including the isotopic impurity \HEF.  This situation changed with the discovery of \SHE\ in silica aerogel at Cornell University~\cite{Por.95} and Northwestern University~\cite{Spr.95} and has led to a number of surprising developments.  Recently, it was found that controlled anisotropy from uniform strain in the aerogel  produces anisotropic quasiparticle scattering to which the relative stability of the superfluid phases is very sensitive, with very different results for positive versus negative strain achieved in silica aerogel by stretching or compressing the framework. This is a central topic of the present review.

%**************************************************************************************************************************
%************************************Figure 1************************************************
%**************************************************************************************************************************
\begin{figure}[t!]
\includegraphics[width=140mm]{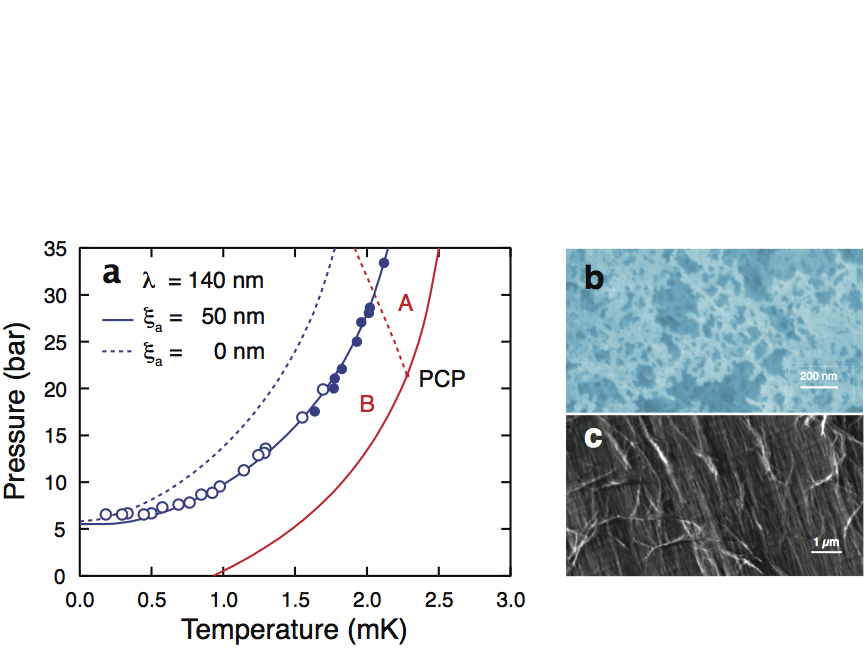}
\caption{\label{PhaseDiagram}  a) Pressure-temperature phase diagram for  \SHE\ in 98\% porous silica aerogel.  The transition from the normal to superfluid in aerogel is the solid blue curve.  It can be compared with the solid red curve for pure \SHE\ where the transition between $A$ and $B$ phases in zero magnetic field is given by the red dashed curve.  The solid-to-liquid melting curve is approximately horizontal near the top of the frame. Open circles are from Cornell University~\cite{Mat.97} torsional oscillator measurements and  closed circles from Northwestern University acoustic measurements.~\cite{Ger.01,Ger.02} The blue curve is a theoretical fit to the data with parameters $\lambda$, the quasiparticle mean-free-path and $\xi_a$, the silica particle-particle correlation length.~\cite{Thu.98,Sau.03} The dashed blue curve is in the absence of these correlations. b) Scanning electron microscope images of a 98\% silica aerogel, and c) a nafen-90 Al$_2$0$_3$ nematic aerogel.~\cite{Asa.15} [figures adapted from  a):~\citenum{Hal.08}; figure: c):Ref.~\citenum{Asa.15}] 
}
\end{figure}

Superfluidity in the pure phases of \HE\ was first observed~\cite{Osh.72,Osh.72b} by  Douglas Osheroff, Robert Richardson, and David Lee who were acknowledged with the Nobel prize
in 1996 for their pioneering work, and to Anthony Leggett in 2003 for the theory he developed~\cite{Leg.72}
in close contact with those experiments. The transition from normal to superfluid is a second order thermodynamic transition, Fig.~\ref{PhaseDiagram}, first detected during Pomeranchuk cooling experiments along the melting curve, a horizontal line at $P =34$\,bar near the top of this figure.  This transition was marked by a change in slope of the measured pressure-time trace.  It was correctly interpreted by Vvedenskii,~\cite{Vve.72} to be attributed to thermal disequilibrium within the liquid-solid mixture, changing abruptly at the superfluid transition where the heat capacity of the liquid increases abruptly by a factor of two.~\cite{Whe.75,Hal.76} Shortly afterward, measurements of the \HE\ nuclear magnetic resonance (NMR) frequency shift were reported~\cite{Osh.72b} that were  precisely accounted for by Leggett's theory.~\cite{Leg.72} Onset of a superfluid fraction was  detected by vibrating wire experiments~\cite{Alv.74} and the propagation of fourth sound.~\cite{Koj.74,Yan.74}  The line of phase transitions that marks  appearance of superfluidity in pure \HE\ is shown by the red curve in Fig. \ref{PhaseDiagram}. The stable  $A$ and $B$ phases in zero external field come together at a poly-critical point (PCP) at the pressure $P = 21$\, bar.  An excellent comprehensive review can be found in the book by Vollhardt and W\"olfle,~\cite{Vol.90}  briefly introduced in the following.

%*********************************************************************************************
%************************************Figure 2************************************************
%*********************************************************************************************
\begin{figure}[t]
\includegraphics[width=120mm]{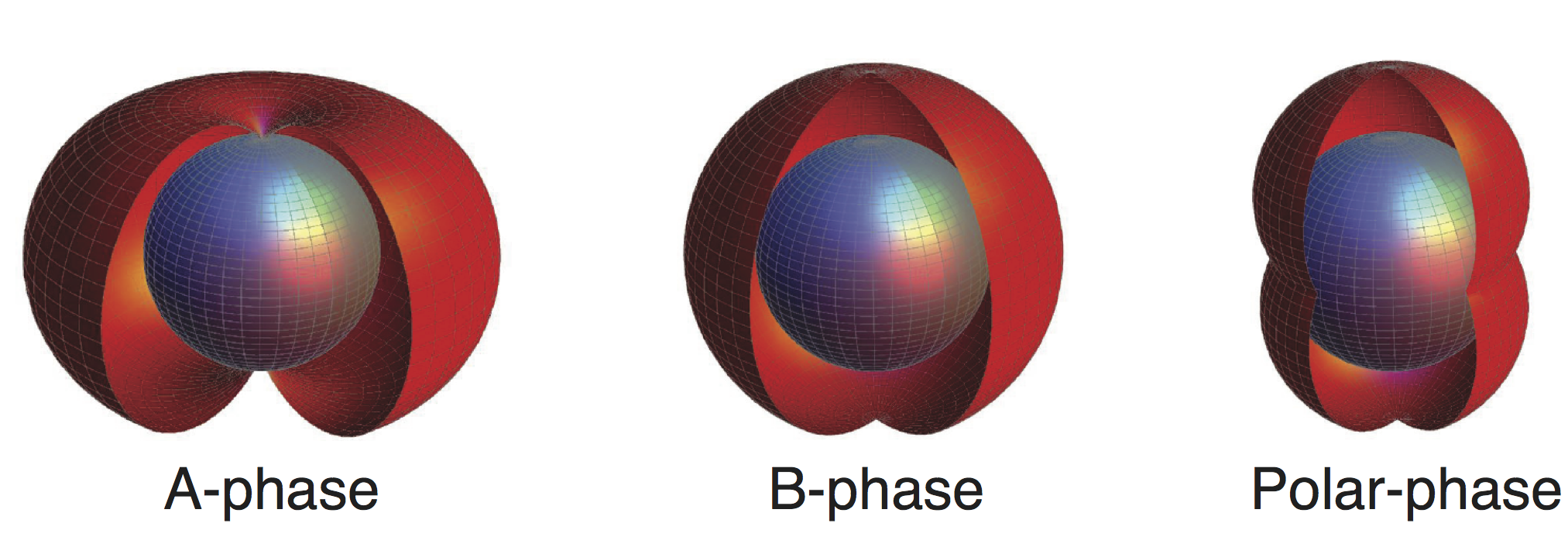}
\caption{\label{Gaps} The energy gap structure as a function of wavevector for the various observed phases of \SHE\ in zero magnetic field showing from left-to-right the $A$ phase (axial state), the $B$ phase (isotropic state), and the polar phase, phases with two-dimensional, three-dimensional, and one-dimensional order parameter structure respecctively. Both $A$ and polar phases are equal-spin-pairing states having the same susceptibility as the normal state.  Only the $A$ phase is chiral and breaks time reversal symmetry.[figure adapted from Ref.~\citenum{Lee.17}]}
\end{figure}

Liquid $^3$He is a degenerate Fermi liquid at temperatures well below the Fermi temperature $T_F \approx 1$\,K where Landau's quasiparticle picture provides a complete phenomenological description.  At temperatures below 2.5\,mK, it transforms to a  $p$-wave, odd-parity superfluid condensate of Cooper pairs with orbital angular momentum $L = 1$ and spin angular momentum $S = 1$. This state was predicted by Anderson and Morel~\cite{And.61} to be the axial state and corresponds to the $A$ phase.  However, the isotropic state  is the more stable phase in weak-coupling theory, as was shown by Balian and Werthamer~\cite{Bal.63} and Vdovin,~\cite{Vdo.63} now known as the $B$ phase.  The $A$, $B$, and polar phases are three states in the $p$-wave manifold~\cite{Vol.90} with order parameters of different symmetry and energy gap structures sketched relative to the Fermi energy, shown as a blue sphere in Fig.~\ref{Gaps} .  The $B$ phase has a broken relative rotation symmetry between spin and orbital degrees of freedom which is responsible for the acoustic Faraday Effect and the unique phenomenon of transverse sound in a fluid.~\cite{Lee.99} The $A$ phase separately breaks both spin and orbital rotation symmetry and time reversal symmetry and is a chiral superfluid.  The $A$ and polar phases are equal spin pairing (ESP) states; their magnetic susceptibility is temperature independent, equal to that of the normal Fermi liquid, and the spin structure of their order parameter is a superposition of the even spin triplet pair states $\ket{\uparrow\uparrow}$ and $\ket{\downarrow\downarrow}$. In contrast, the $B$ phase is a non-ESP state with a strongly temperature dependent susceptibility.  Its spin order parameter is a combination of, $\ket{\uparrow\uparrow}$ and  $\ket{\downarrow\downarrow}$, and $\ket{\uparrow\downarrow+\downarrow\uparrow}/\sqrt{2}$ spin pairs and the transition between $A$ and $B$ phases is first order.  The polar phase does not break time reversal symmetry and does not exist in pure \HE.  However, it was predicted by Aoyama and Ikeda to occur in  anisotropic aerogel.~\cite{Aoy.06} It was recently observed from NMR frequency shift measurements by Dmitriev {\it et al.}~\cite{Dmi.15} in a nematically ordered Al$_2$O$_3$ aerogel, Fig.~\ref{PhaseDiagram}\,c).  The polar phase has a one-dimensional order parameter structure as compared with the two-dimensional $A$ phase and three-dimensional $B$ phase.  As with superconductors, the important length scale in \SHE\ is the coherence length, a measure of the Cooper pair size, $\xi(P) = \hbar v_F/2\pi k_B T_c$, that varies from 77 nm at $P = 0$ to 16 nm at $P = 34$\,bar with $v_F$ the Fermi velocity and $T_c$ the transition temperature.\\

%*********************************************************************************************
%************************************Figure 3************************************************
%*********************************************************************************************
\begin{figure}[b!]
\includegraphics[width=140mm]{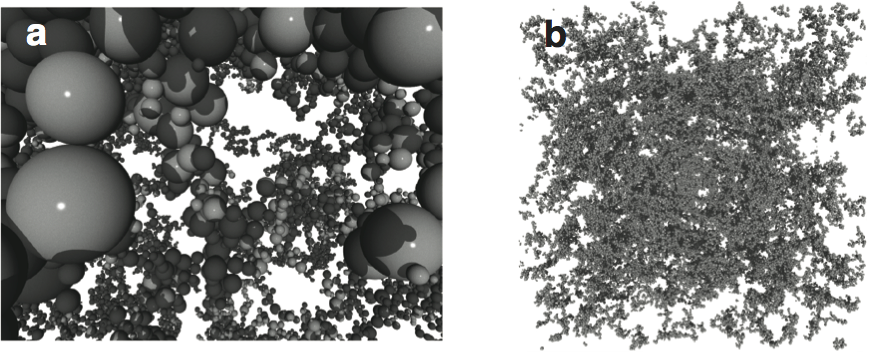}
\caption{\label{DLCA}
a) Perspective view from numerical simulation of a 98\% porosity gel structure grown using a  diffusion limited cluster aggregation algorithm and b) the 3D simulation.~\cite{Haa.00,Haa.01}  The 3\,nm particles assemble in strands that are spatially correlated over distances $\xi_a$ of order 30\,nm.[figure adapted from Ref.~\citenum{Li.13}]
}
\end{figure}

\subsection*{Silica Aerogel}

Aerogels are highly porous bodies that can be produced from a wide range of  materials including silica, carbon, and alumina among others.  Silica aerogels up to 99.5\% porosity can be formed from a base-catalyzed synthesis of silica nanoparticles approximately 3\,nm in diameter.~\cite{Fri.88} Gelation is initiated from
tetramethylorthosilicate and the particles aggregate to form a fractal structure shown in the scanning electron microscope image in Fig.~\ref{PhaseDiagram}\,b) and numerically simulated in Fig.~\ref{DLCA}.~\cite{Haa.00}  Both give a fractal dimension of $\approx 1.7$ for a 98\% porosity aerogel, the material most commonly used for the study of effects of quenched disorder on \SHE.~\cite{Pol.08,Hal.08} 
The wet gel is dried at a supercritical pressure for the methanol solvent using  a high pressure autoclave to avoid collapse of the microstructure from capillary forces at the liquid-gas interface. The resulting material is air stable and
hydrophobic.  Pollanen {\it et al.}~\cite{Pol.08} found that a `one-step' method~\cite{Tei.85} with relatively small amounts of catalyst can produce very uniform structures. This procedure is important to obtain high quality samples and was greatly facilitated by characterization using optical birefringence~\cite{Pol.08,Bhu.99} and with small angle X-ray scattering (SAXS).~\cite{Por.99,Pol.08}  Haard {\it et al.} used a diffusion-limited-cluster-aggregation algorithm  to simulate 98\% porosity gel structures, Fig.~\ref{DLCA}, indicating that density correlations persist below the upper fractal cutoff in the particle distribution $\simeq 100$\,nm.~\cite{Haa.00}  They  identified a correlation length, $\xi_a \approx 30$\,nm, as the typical distance between silica strands.~\cite{Haa.00,Haa.01}  At longer length scales the aerogel particle-particle correlation length is a measure of the more open voids.  The simulated structure has a geometric mean-free-path, $\lambda$  $\simeq 200$\,nm, defined as the average length of a straight line trajectory terminating at  aerogel surfaces and  corresponds well to the transport mean-free-path resulting from elastic scattering of the $^{3}$He quasiparticles from the aerogel in the normal Fermi liquid. Indeed, this is borne out by analysis of  measurements in the normal Fermi liquid including thermal conductivity,~\cite{Sha.03} and spin diffusion,~\cite{Sau.05} giving transport mean-free-paths, $\lambda\approx$ 150 to 180\,nm.  Silica aerogel is quite reversibly compliant to compressive (negative) strain up to $\simeq30$\%.  Similar levels of positive strain, called stretching, can be produced during  stages of growth and supercritical drying.~\cite{Pol.08}  Furthermore, the optical birefringence signal provides a simple but quantitative assessment of the degree of anisotropy imposed on the aerogel by either positive or negative strain,~\cite{Zim.13,Li.15} confirmed with measurements of the mean free path using methanol gas diffusion.~\cite{Lee.14} For nafen aerogels the anisotropy of the mean-free-path  was measured by spin diffusion of \HE\ in the normal liquid.~\cite{Dmi.15b} Aerogel samples are considered to be isotropic if they have minimal strain based on these characterizations and for which quasiparticle scattering is isotropic.

\subsection*{ Superfluid \HE\ in Aerogel}
\begin{figure}[b]
\includegraphics[width=160mm]{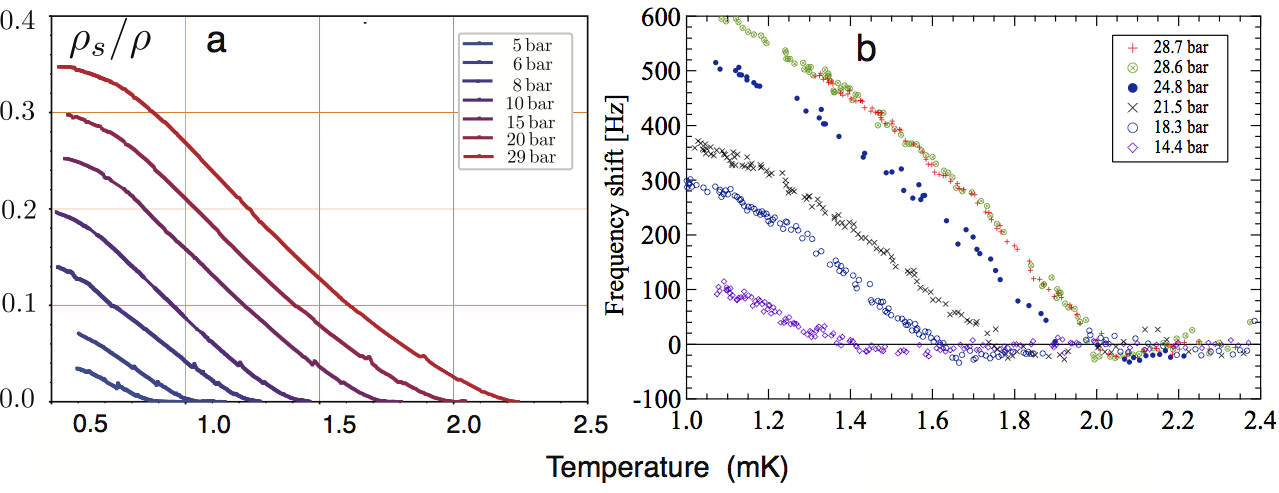}
\caption{\label{Discovery} The first measurements of \SHE\ in silica aerogel. a) Superfluid fraction from torsional oscillator experiments at Cornell University.~\cite{Por.95}  The superfluid fraction is reduced with respect to that in pure \SHE\ approaching zero at low pressure. b) NMR frequency shift $\Delta\omega$ proportional to the square of the amplitude of the order parameter from Northwestern University,~\cite{Spr.95} showing a  reduction in magnitude  from pure \HE, as is the case for $\rho_s/\rho$. [figure adapted from a):Ref.\citenum{Por.95};  b):Ref.\citenum{Spr.95}]
}
\end{figure}

The discovery that \SHE\ survives in the aerogel environment~\cite{Por.95,Spr.95} was a surprise given that aerogel has a fractal distribution of length scales and the fact that any form of \HE\ quasiparticle scattering, such as from the distributed structure of aerogel, breaks Cooper pairs. However, the important impurity length scale is not the fractal cutoff, nor the silica particle-particle correlation length, $\xi_a$;  rather, it is the much larger quasiparticle mean-free-path $\lambda$.  The first indications of a superfluid state in aerogel are from measurements of the superfluid fraction and NMR frequency shift  shown in Fig.~\ref{Discovery}.

%*********************************************************************************************
%************************************Figure 4************************************************
%*********************************************************************************************

\begin{figure}[t!]
\hspace*{-60mm}
\includegraphics[width=200mm]{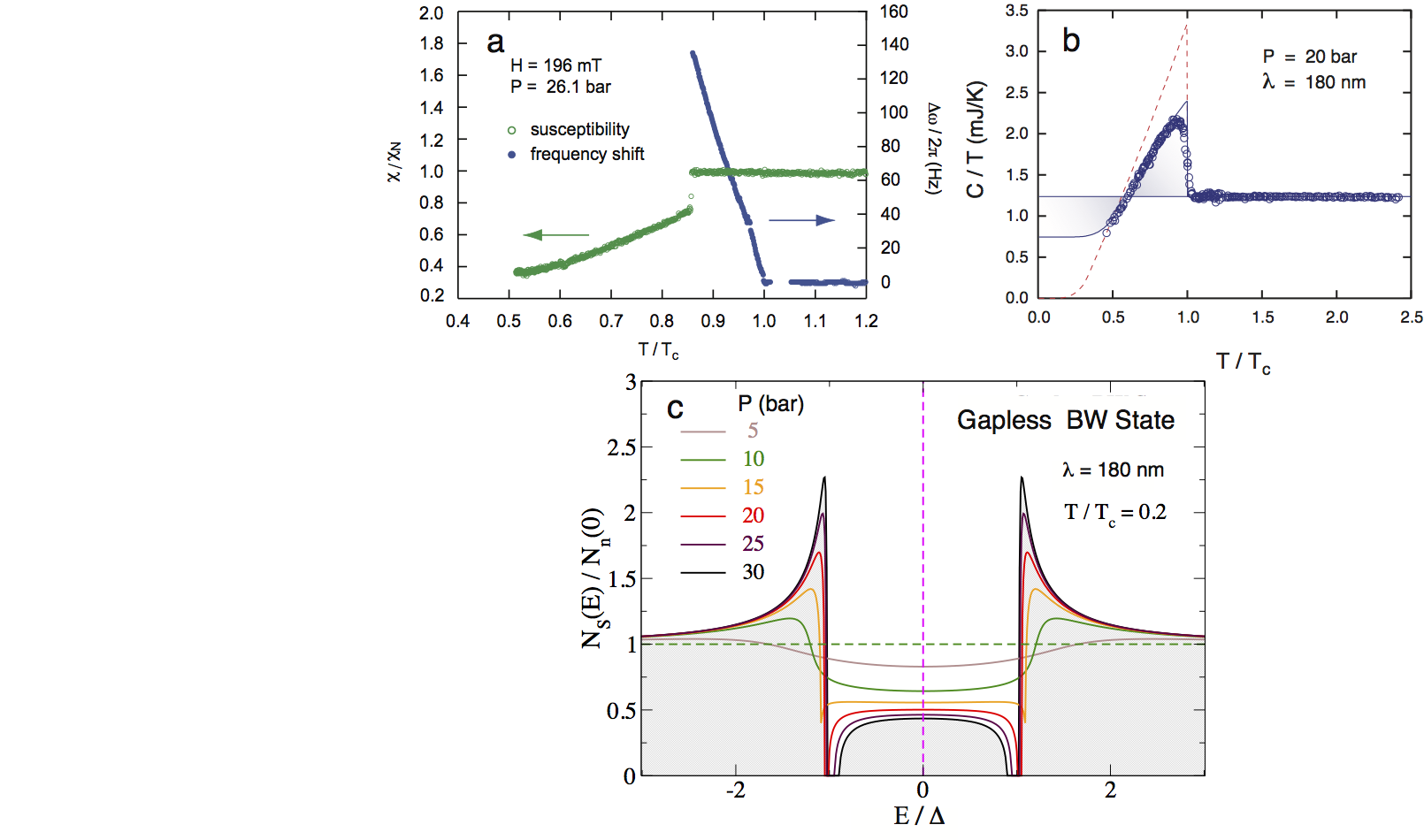}
\caption{\label{SharpTransitionsandDOS}
a) Sharp transitions for \SHE\ in a 98\% isotropic aerogel sample from susceptibility, and NMR frequency shift taken on warming.~\cite{Pol.11} b) Heat capacity  as a function of temperature compared to pure \SHE\ (dashed red curve).~\cite{Cho.04}  The low temperature limit of the heat capacity must be linear in temperature to conserve entropy in agreement with theory.~\cite{Ali.11,Sauls_PC}  c) Calculated density of states  for the $B$ phase in aerogel with a mean-free-path of $\lambda=180$\,nm showing gapless behavior at the Fermi energy where $\Delta$ is the amplitude of the order parameter.~\cite{Sha.03,Sha.03b} [figure adapted from a):~\citenum{Pol.11};  b):Ref.~\citenum{Cho.04}; c):Ref.~\citenum{Sha.03b}]
}
\end{figure}

This prompted the formulation of a Ginzburg-Landau description of \SHE\ in aerogel by Thuneberg {\it et al.},~\cite{Thu.98} that included strong coupling in the pairing interaction and was extended by Sauls and Sharma~\cite{Sau.03} to include the silica particle-particle correlation length.  The predicted phase diagram accurately fits the experiment with these two parameters, mean-free-path and correlation length, given by the solid blue curve in Fig.~\ref{PhaseDiagram}.  The low pressure region of the phase diagram is determined by a pair breaking parameter given by the ratio of the pure superfluid coherence length to the mean-free-path  where  correlation effects are absent and a critical pressure appears, not present in pure \HE. This result is exactly analogous to the critical value of the pair breaking parameter, $\xi/\lambda$, in the Abrikosov Gorkov theory for superconductors.~\cite{Abr.61} On the other hand, at high pressures the effects of the correlation length are very much in evidence.  Not only is the observed suppression of the transition temperature modest in this regime, but the inhomogeneous broadening of the superfluid transition can be very small with the thermodynamic transitions between phases very sharply defined.  With high quality samples, this can be of the order of  $2 \,\mu$K, Fig.~\ref{SharpTransitionsandDOS}.~\cite{Pol.11} From the early experiments, such as in this figure, it was natural to think that the high temperature ESP phase in a magnetic field was analogous to the $A$ phase of pure \SHE, while the phase with a temperature dependent susceptibility was like the pure $B$ phase.  Consequently, they were often referred to as $A$-like and $B$-like phases until  a firm identification of the symmetry of these phases came from  NMR experiments that are discussed later.

%********************************************************************************************
%************************************Figure 5**********************************************
%********************************************************************************************

\subsection*{Superfluid Fraction}

The extreme resolution of a high-Q torsional oscillator is ideal for measurement of the superfluid state and gave the  first evidence for the existence of \SHE\ in aerogel.~\cite{Por.95}  The oscillator consists of a disk containing the helium-aerogel  sample, perpendicular to its torsion rod,  for which a change in period at resonance can be precisely measured.  According to the two-fluid model, a superfluid is the superposition of a normal fluid component, viscously clamped to the porous structure, plus an inviscid superfluid component that does
not contribute to the moment of inertia. Consequently, at the onset of superfluidity there is a
sharp decrease of the oscillation period. The loss of inertia can be quantitatively
interpreted in terms of the superfluid fraction, shown in Fig.~\ref{Discovery}\,a).
However, in contrast with
pure superfluid $^3$He, the superfluid fraction is significantly less than unity, and the more so at low pressure. The suppression in the superfluid fraction varies from about 40\% at high pressure to zero at the critical pressure shown in Fig.~\ref{PhaseDiagram}.  

\subsection*{NMR Frequency Shift}
The  `smoking gun' for superfluidity in pure \HE\ came from NMR measurements and Leggett's interpretation of them.~\cite{Leg.75} There are three important NMR measurements that are manifestations of superfluid order.  First, the magnetic susceptibility is proportional to the integral of the NMR frequency spectrum that easily distinguishes an ESP from a non-ESP superfluid state. Second, the NMR spectrum shifts as a function of temperature by an amount $\Delta\omega$ from the Larmor frequency, {\it i.e.} the position of the resonance in the normal fluid, and exhibits an abrupt onset on cooling.  This is clearly observed for \SHE\ in aerogel in Fig.~\ref{Discovery}\,b).  Third, the nuclear spin dynamics embodied in the Leggett equations~\cite{Leg.75,Vol.90} give rise to a dependence of $\Delta\omega$ on the magnitude of the pulsed  NMR excitation that is specific to each superfluid state.  The excitation amplitude is best represented as the tip angle, $\beta$, of the local nuclear magnetization away from the external magnetic field.  Together these three measurements can provide a clear identification of the superfluid state.  Most importantly, the magnitude of the frequency shift is directly related to the square of the amplitude of the order parameter shown for the different phases in Fig.~\ref{Gaps}.  

The order parameters of the three superfluid phases, with maximum gap amplitudes, $\Delta_A$, $\Delta_B$, and $\Delta_P$ given in Fig.\ref{Gaps}, are for $A$ (axial state), $B$ (isotropic state), and $P$ (polar state) phases expressed as:~\cite{Lee.17,Vol.90}
\begin{eqnarray}
\begin{aligned}
A^A_{\mu j} & =\Delta_A\hat{d}_{\mu}(\hat{m_1}_j+i\hat{m_2}_j)e^{i\phi}\nonumber\\
A^B_{\mu j} & =\Delta_BR_{\mu j}e^{i\phi}\nonumber\\
A^P_{\mu j} & =\Delta_P\hat{d}_{\mu}\hat{p}_je^{i\phi}\nonumber,\\
\end{aligned}
\end{eqnarray}
\noindent
where the order parameter unit vectors are $\hat{d}\perp\hat{s}$ in spin space, with orthogonal orbital vectors,  $\hat{m_1}$, and $\hat{m_2}$.  The direction of angular momentum in the $A$ phase is specified by $\hat{\ell} = \hat{m_1}\times\hat{m_2}$ along the axis of the nodes of the energy gap; and the orbit and spin coordinates are $j$ and $\mu$.  The $B$ phase  order parameter is specified by a rotation matrix $R_{\mu j}$ which locks spin $\hat{s}$ and orbit $\hat{\ell}$ directions by a rotation through the Leggett angle $\theta_L=104^\circ$ about an arbitrary axis $\hat{n}$.  The polar axis in the $P$ phase is $\hat{p}$. However, external influences  from magnetic field, walls, flow, and aerogel anisotropy can influence the direction of the spin, and orbital directors, $\hat{d}$, $\hat{s}$, $\hat{l}$, and $\hat{n}$ in order to minimize the total free energy including condensation energy,  Zeeman energy and dipole energies, the last being the weakest by far.  For example, the angular momentum $\hat{\ell}$ is perpendicular to a wall for all phases; similarly, for anisotropic aerogel,  $\hat{\ell}$ must be either aligned parallel or perpendicular to the anisotropy axis and both situations occur. These external fields are  sometimes in competition with each other to determine the texture of the vector order parameter with possible ensuing topological defects.~\cite{Vol.90}  Nonetheless, there are two well-defined stable texture configurations which are relevant to the identification of each superfluid phase in aerogel and can be easily detected. These are the dipole-locked and the dipole-unlocked configurations corresponding respectively to minimum and maximum  dipole energy for the accessible equilibrium phases. 

For the (undistorted) $B$ phase in a magnetic field, $H$, greater than the dipole field of $\approx 3$\,mT,  the dipole-locked case requires that $\hat{H},\hat{s}$ and $\hat{\ell}$, and therefore $\hat{n}$, all be parallel, often referred to as the Brinkmann-Smith configuration.~\cite{Bri.75} This holds if there are no other fields competing with the external field. The dipole-unlocked case corresponds to $\hat{H}\parallel\hat{s}$ with both perpendicular to $\hat{\ell}$, and with $\hat{n}$ at an angle of cos$^{-1}\sqrt{1/5}=63^{\circ}$ relative to each, for example near a surface with field in-plane.  The dependence of $\Delta\omega$ on tip angle $\beta$ in both cases is shown in Fig.~\ref{5/2} and compared with the measurements for an isotropic aerogel.~\cite{Pol.11} For the (undistorted) $A$ phase the dipole-locked case with minimum dipole energy has $\hat{\ell}\parallel \hat{d}$ and $\hat{\ell}\perp\hat{s}$.  For the dipole-unlocked configuration it is $\hat{\ell}\perp\hat{d}$.

The NMR frequency shift for small tip angles, $\beta$, is an excellent indication of the symmetry of the order parameter for which direct comparisons of $\Delta\omega$ can be made between the dipole-unlocked case of the $B$ phase and the dipole-locked case for the $A$ phase based on  the symmetry of the two states.~\cite{Zim.18}  A similar situation holds for comparison of the dipole-locked shifts for $P$ and $A$ phases:~\cite{Zim.18}

\begin{eqnarray}
\frac{\Delta\omega_B}{\Delta\omega_A}&=&2\frac{\Delta C_{B}}{\Delta C_{A}}\frac{\chi_{A}}{\chi_{B}} = 3\frac{\Delta^2_B}{\Delta^2_A}\frac{\chi_A}{\chi_B}\nonumber\\
\frac{\Delta\omega_P}{\Delta\omega_A}&=&2\frac{\Delta C_{P}}{\Delta C_{A}} = \frac{\Delta^2_P}{\Delta^2_A},
\label{ratios}
\end{eqnarray}

\noindent
 expressed in terms of the maximum gap amplitudes; or alternatively in terms of the heat capacity jumps at $T_c$ including strong coupling effects for both cases.  In the weak coupling approximation for pure \HE\ the two ratios are respectively: 12/5 and 4/3.

\subsection*{Isotropic Silica Aerogel: Identification of the Superfluid Phases by NMR}

For \SHE\ in aerogel NMR measurements of the frequency shift $\Delta \omega$ at small tip angle  $\beta\sim10^{\circ}$ shown in Fig.~\ref{Discovery}\,b) and \ref{SharpTransitionsandDOS}\,a),~\cite{Spr.95,Pol.11} have a sharp onset  very similar to that for pure \SHE. However, there is a reduction in  magnitude of the shift, and therefore the amplitude of the order parameter, as compared with pure \SHE\ similar to the  superfluid fraction Fig.~\ref{Discovery}\,a), and the heat capacity jump,~\cite{Cho.04} Fig.~\ref{SharpTransitionsandDOS}\,b). Together these results provide a consistent picture of pair breaking from quasiparticle scattering from aerogel surfaces and suppression of the magnitude of the order parameter.~\cite{Thu.98, Ali.11} 

In an isotropic aerogel the spin and orbital directions of the superfluid are unconstrained by the medium allowing a `dipole-locked' configuration with uniform texture that gives the minimum possible dipole energy. A class of such samples that are strain-free on the sub-micron scale have been grown and characterized by Pollanen {\it et al.}~\cite{Pol.08}   The NMR frequency shifts, $\Delta \omega (T)$, are well-resolved  allowing an unambiguous identification of the superfluid states and a measure of the temperature dependent energy gap, $\Delta \omega (T) \propto \Delta^2(T)$.

\begin{figure}[b!]
\includegraphics[width=160mm]{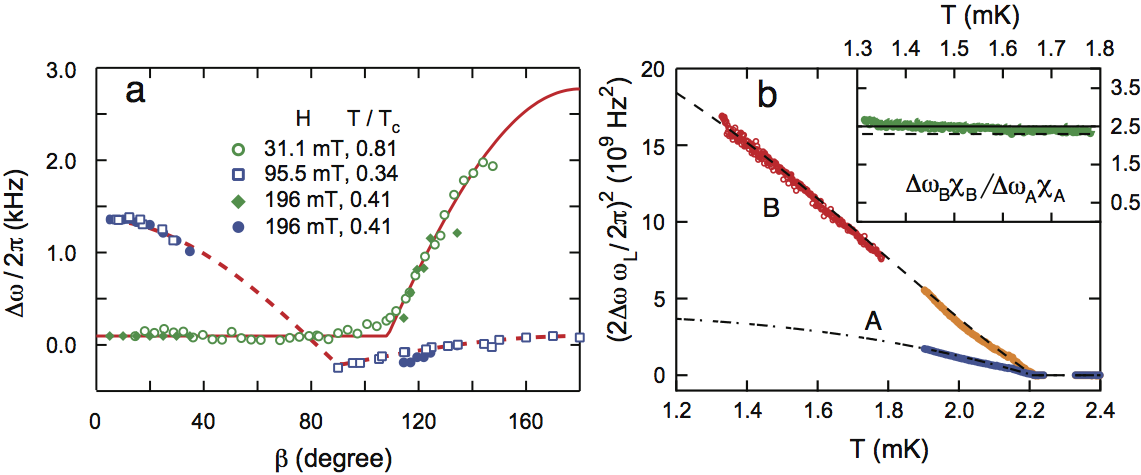}
\caption{\label{5/2}
NMR frequency shift $\Delta\omega$ in an isotropic aerogel at $P=26$ bar scaled to a common field of $H=196$ mT.~\cite{Pol.11}  a) Tip angle dependence for the dipole-locked  (green) and dipole-unlocked (blue) configurations are compared with theory, solid and dashed red curves. b) Temperature dependence of $\Delta\omega$ in the $B$ phase (red)  and $A$ phase (blue).  The orange data are scaled from the $A$ phase data using $5/2$ from Eq.~\ref{ratios} and the measured magnetic susceptibility, consistent with extrapolation to $T_c$ of $\Delta\omega$ from the $B$ phase (dashed line). The dashed-dot curve is the measured temperature dependence for the $A$ phase.~\cite{Bau.04} Inset: The ratios of $B$ to $A$ phase data are equal to 5/2 which on symmetry grounds corresponds to isotropic and axial states, Eq.~\ref{ratios}.
 [figure adapted from: Ref.~\citenum{Pol.11}]
}
\end{figure}

Using isotropic aerogel, Pollanen {\it et al.}\cite{Pol.11} found well-separated dipole-locked and dipole-unlocked parts of their NMR spectrum in the non-ESP phase, the latter from the influence of sample walls that show up at low temperatures. These frequency shifts were identified from their tip angle dependence closely following theoretical expectations for the $B$ phase and are shown in Fig.~\ref{5/2}\,a). The measurements of $\Delta\omega$($\beta$), together with a temperature dependent susceptibility, Fig.~\ref{SharpTransitionsandDOS}\,a), confirm that this low temperature phase is indeed the isotropic state, a suppressed version of the pure $B$ phase. The high temperature ESP phase in Fig.~\ref{SharpTransitionsandDOS}\,a) results, at least in part, from the application of a magnetic field which destabilizes the $B$ phase; however, this gives no information about the symmetry of the superfluid state of that phase.  To determine symmetry we compare $\Delta\omega$ at small tip angles as suggested by Eq.~\ref{ratios} and shown in Fig.~\ref{5/2}\,b) with the $B$ phase finding exactly the ratio of 5/2.   This is also shown in the figure's inset, consistent with the two phases being respectively isotropic and axial states, {\it i.e.}  identifying the ESP phase as the $A$ phase.

%*********************************************************************************************
%************************************Figure 6************************************************
%*********************************************************************************************

Concerning phase diagrams and the relative stability of $A$ and $B$ phases, Gervais {\it et al.}~\cite{Ger.01,Ger.02} reported from acoustic impedance measurements that the equilibrium superfluid state in an isotropic aerogel in zero magnetic field was the $B$ phase throughout the entire pressure-temperature phase diagram, Fig.~\ref{AnisotropicPDs}\,a). This was confirmed in a well-characterized isotropic aerogel from NMR measurements of $T_{AB}$, from magnetic susceptibility and $\Delta\omega$ extrapolated to zero field,~\cite{Pol.11,Pol.12} Fig.~\ref{AnisotropicPDs}\,b) and d).  Both the acoustic and NMR data were taken on warming in order to determine the equilibrium transition.  On cooling, a metastable supercooled $A$ phase always appears, even in the absence of an applied field.~\cite{Ger.02} Presumably there is a narrow unobserved region just below the superfluid transition where the $A$ phase is stable that leads to its supercooling as low as 0.83 $T_{AB}$. However, the  origin for the nucleation of this metastable $A$ phase immediately below $T_c$,  is unknown and the supercooling is quite different from  pure \SHE\ especially below the PCP.  

The fact that in the absence of magnetic field the $B$ phase  is more stable than the $A$ phase in isotropic aerogel is consistent with the Ginzburg-Landau theory of Thuneberg {\it et al.},~\cite{Thu.98} that isotropic quasiparticle scattering favors an isotropic superfluid state. The stability of the $A$ phase in pure \HE\ at high pressure, Fig.~\ref{PhaseDiagram}, is a consequence of strong coupling in the pairing interaction.  Apparently, this is overshadowed by the influence of isotropic  quasiparticle scattering.  However, in a magnetic field, the $A$ phase becomes more stable owing to its different Zeeman energy relative to the $B$ phase.  The transition between $A$ and $B$ phases in a magnetic field is first order where $T_{AB}$ depends linearly on the square of the magnetic field in the Ginzburg-Landau regime near $T_c$, clearly evident in Fig.~\ref{AnisotropicPDs}\,b) and d). 

\subsection*{Gapless Superfluidity}

Heat transport~\cite{Fis.03} and heat capacity measurements,~\cite{Cho.04} Fig.~\ref{SharpTransitionsandDOS}\,b), indicate that their low-temperature limits
for  \SHE\ in aerogel are linear in temperature, consistent
with  a significant density of gapless fermionic excitations near the Fermi level,~\cite{Sha.03,Ali.11,Sauls_PC} Fig.~\ref{SharpTransitionsandDOS}\,c). The third law of thermodynamics requires that the entropy of both the normal and
superfluid phases vanish at zero temperature.  This constraint on the heat capacity forces the two shaded regions in Fig.~\ref{SharpTransitionsandDOS}\,b) to have equal areas resulting in the extrapolation of the data for $C/T$ to low temperatures being non-zero.  Consequently, the heat capacity
must be linear in $T$ at low temperatures consistent with  theory.~\cite{Ali.11,Sauls_PC}  By comparison, pure  \SHE-$B$ is fully gapped over the entire Fermi surface and its heat capacity becomes exponentially small at low temperatures.  A similar argument holds for the thermal conductivity.~\cite{Sha.03} The evidence is compelling from both of these thermal experiments that liquid \HE\ in aerogel is a gapless superfluid  and that this holds for both $A$ and $B$ phases based on calculation of their density of states, shown only for the $B$ phase in Fig.~\ref{SharpTransitionsandDOS}\,c).~\cite{Sha.03,Sha.03b}

\subsection*{Superfluid Glass}
The $A$ phase angular momentum for pure, unconstrained \SHE\ can assume any orientation and maintain the minimum dipole energy.  Volovik pointed out~\cite{Vol.96} that for \HE-$A$ in isotropic aerogel, this continuous symmetry of the ordered state will be sensitive to even an arbitrarily small amount of disorder precluding long range orientational order of $\hat \ell$. The arguments were first developed by Larkin~\cite{Lar.70} to understand the effect of disorder on the vortex state in superconductors and by Imry and Ma~\cite{Imr.75} from a general theoretical perspective for any vector order parameter with a continuous symmetry.  Thus, \HE-$A$  should be an orbital glass in the presence of the quenched disorder of aerogel, with the caveat that the length scale of this disorder be less than the dipole length, $\simeq8$\,microns,~\cite{Vol.90} and that there be no macroscopic strain or anisotropic inhomogeneity in the aerogel on greater length scales than this that mask the phenomenon.  This three-dimensional LIM effect would be characterized by no observable NMR frequency shift and no contribution to the NMR linewidth beyond that of the normal state.~\cite{Li.13}   This means that superfluid order in a true LIM state would not be detectable by NMR.  However, various manifestations of a LIM state have been reported,~\cite{Elb.08,Dmi.10} but in some cases it was  completely absent, Fig~\ref{SharpTransitionsandDOS}.~\cite{Pol.11} It was eventually realized that  for $A$ phase data taken on warming from the $B$ phase there must be a broken rotational symmetry possibly established at the $A$-$B$ interface as it progresses through the superfluid creating a metastable state of orbital nematic order.  This idea was confirmed by Li {\it et al.}~\cite{Li.13} who found that on cooling from the normal state there were no NMR signatures  of a superfluid in the linewidth or frequency shift, consistent with a LIM state where the nematically ordered $A$ phase had  been observed on warming from the $B$ phase. This provided an unambiguous signature of the orbital glass state in \SHE\ in the presence of random disorder.

\subsection*{Anisotropic Silica Aerogel and New Superfluid Phases}

\begin{figure}[!]
\includegraphics[width=120mm]{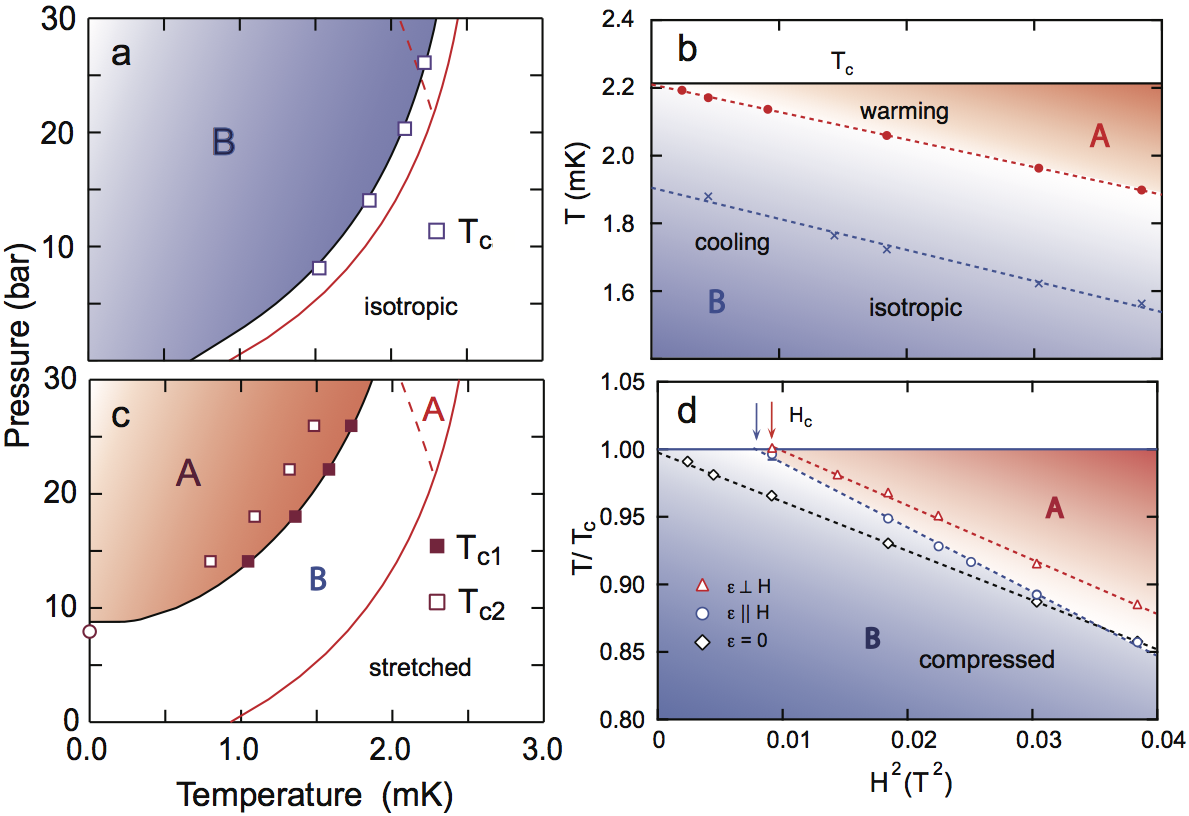}
\caption{\label{AnisotropicPDs} a) and b) Phase diagrams for  \SHE\ in isotropic silica aerogel. a) The pressure-temperature phase diagram from acoustics~\cite{Ger.01,Ger.02} and from NMR extrapolated to $H=0$ for isotropic aerogel. The $B$ phase is stable at all pressures, different from pure \HE.~\cite{Pol.11}  Note that this sample has less impurity scattering than that shown in Fig.~\ref{PhaseDiagram}. b) The temperature-magnetic field phase diagrams at  $P = 26$\,bar for the equilibrium transition from  $B$ to $A$ on warming (red), and for the metastable supercooled $A$ phase (blue), as a function of $H^2$. The $A$ phase supercools at all fields, consistent with acoustic measurements at different pressures,~\cite{Ger.01,Ger.02}  and the phase in the region between red and blue shading depends on this history. c) The pressure-temperature phase diagram for a  stretched aerogel with strain $\epsilon = 14\%$ from NMR measurements over the same range of $H$ as in b). There are two different ESP states independent of pressure.~\cite{Pol.12,Li.14b}  The higher temperature phase is the $A$ phase which appears  from the normal state at $T_{c1}$ with angular momentum, $\hat{\ell}\parallel\hat{\epsilon}$.  The lower temperature phase at $T_{c2}$ is  the $A$ phase but with $\hat{\ell}\perp\hat{\epsilon}$. d) The compression of the isotropic aerogel~Fig.\ref{SharpTransitionsandDOS} at $P=26$\,bar with strain $\epsilon = -20\%$ stabilizes a polar-distorted $B$ phase.  On warming, that phase  competes favorably with the $A$ phase in a magnetic field, creating a tri-critical point with a critical field, $H_c$.~\cite{Li.14b,Li.15}  The phase in the region between red and blue shading depends on the sample.[figure adapted from: a) and c):Ref.~\citenum{Pol.12}; b) and d):Ref.~\citenum{Li.13}]
}
\end{figure}

The phase diagrams of the superfluid in anisotropic silica aerogel  are very different compared with isotropic aerogel,~\cite{Pol.11} even to the extent that new phases appear,~Fig.\ref{AnisotropicPDs}\,c) and d).~\cite{Pol.12,Li.14b,Li.15}   Uniform uniaxial anisotropy has been produced in cylindrical samples of nominally 98\% porosity  with  anisotropy and uniformity quantified by optical birefringence.~\cite{Pol.08,Li.14b,Li.15} The optical measurement was confirmed by spin diffusion experiments of the ballistic mean-free-path anisotropy in a stretched aerogel.~\cite{Lee.14} Stretching is achieved using more than the usual amount of catalyst for gelation to promote radial shrinkage during supercritical drying.~\cite{Pol.08,Pol.12} Similarly, uniform anisotropy can be obtained by physical compression of the aerogel up to $\sim$\,-30\% negative strain.~\cite{Li.15} In the research that followed  it became clear that the \SHE\ angular momentum axis, $\hat \ell$, is oriented by uniaxial strain. The first indications of the influence on \SHE\ from anisotropy in the quasiparticle scattering came from acoustic impedance experiments by Davis {\it et al.}, for both stretched~\cite{Dav.06} and compressed~\cite{Dav.08b} silica aerogels.

%********************************************************************************************
%************************************Figure 7**********************************************
%********************************************************************************************

  %********************************************************************************************
  %************************************Figure 8**********************************************
%********************************************************************************************
\begin{figure}[b!]
\includegraphics[width=120mm]{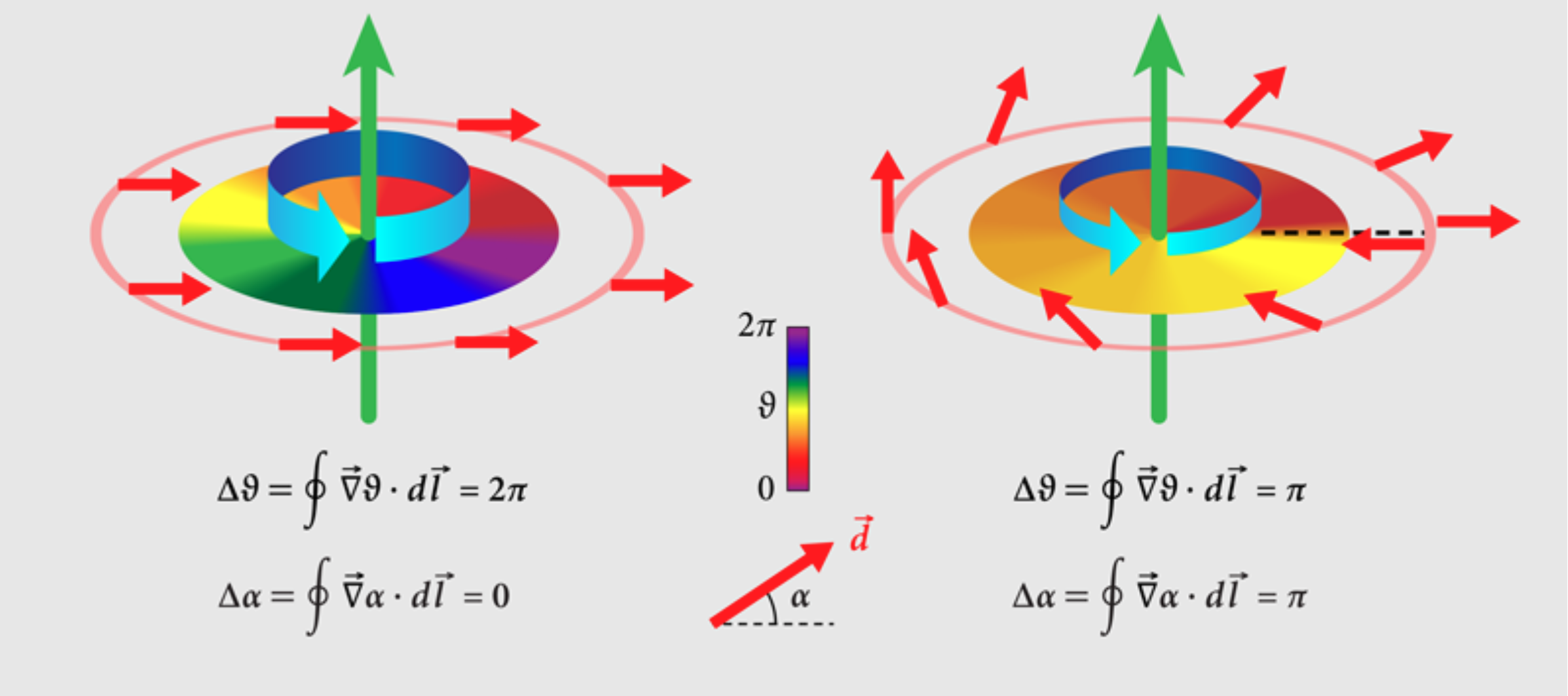}
\caption{\label{HQV} Sketch of the phases of the orbital, $\Delta \theta$, and spin, $\Delta \alpha$, components of the order parameter  showing phase winding around the core of a half-quantum vortex on the right.  On the left the vortex has a full quantum of circulation, $h/2m$.~\cite{Sau.16}\,[figure adapted from: Ref.~\citenum{Sau.16}]
}
\end{figure}

Using NMR, Pollanen {\it et al.}~\cite{Pol.12} showed that the anisotropy introduced by stretching with strain $\epsilon = 14\%$, has the significant effect of stabilizing the $A$ phase throughout the entire pressure-temperature phase diagram, Fig.~\ref{AnisotropicPDs}\,c).  This phase was predicted to be more stable than the $B$ phase since anisotropic quasiparticle scattering favors anisotropic superfluid states.~\cite{Thu.98,Sau.13}   In fact a second  transition appears at lower temperatures, shown as open squares in this figure,  predicted by Sauls to be a biaxial chiral state.~\cite{Sau.13} NMR experiments~\cite{Li.14a,Li.14b} with magnetic fields oriented parallel and perpendicular to the anisotropy axis show that the higher temperature ESP state is indeed the $A$ phase  with angular momentum constrained to be along the strain axis, $\hat{\ell}\parallel\hat{\epsilon}$,  consistent with theory.~\cite{Sau.13} However, in the lower temperature phase the angular momentum abruptly reorients by $90^{\circ}$.~\cite{Pol.12,Li.14c}  This new phase is also a chiral phase although the mechanism for the order parameter reorientation is not yet established.  A different theory~\cite{Vol.06} predicts that $\hat{\ell}\perp\hat{\epsilon}$ would be the equilibrium state, but this appears to be the case only in the lower temperature phase.

The phase diagram for \SHE\ in anisotropic aerogel is quite different for anisotropy introduced with uniform axial compression, {\it i.e.} negative strain,  Fig.~\ref{AnisotropicPDs}\,d).  In low magnetic fields $H\sim100$\,mT the $B$ phase is sufficiently distorted by the compressed aerogel that it is more stable than the $A$ phase in contrast with isotropic aerogel.  The resulting critical field marks a tri-critical point  between the normal phase,  the $A$ phase, and the polar distorted $B$ phase. The polar distortion of the order parameter in the $B$ phase is independently indicated by anomalously large NMR frequency shifts.~\cite{Li.15}  As might be expected, the square of the critical field was found to be proportional to the strain induced anisotropy of the aerogel.~\cite{Li.15} Since NMR measurements have not been made at very low magnetic field close to $T_c$, this region of the phase diagram for compressed aerogels has only been explored by torsional oscillator techniques where a new phase was reported.~\cite{Ben.11}  The superfluid fraction in this  phase appears to be smaller than in the distorted $B$ phase indicating that the angular momentum axis might be oriented perpendicular to the strain as  expected for a polar phase with the polar axis and  anisotropy axis aligned.  Confirmation of the existence of a polar phase in compressed silica aerogel will require further investigation.

\subsection*{Nematic Alumina Aerogel and the Polar Phase}

A class of  nematic aerogels has been produced from the growth of oriented Al$_2$O$_3$ strands.   One of these materials, nafen-243 Fig.~\ref{PhaseDiagram}\,c), has been used to impose  a high level of anisotropy on quasiparticle scattering in the superfluid.~\cite{Dmi.15}  The degree of anisotropy in the mean-free-path was determined from measurement of the spin-diffusion coefficient in the normal Fermi liquid reported to be a factor of 8 for nafen-243.~\cite{Dmi.15b}  Both NMR frequency shift~\cite{Dmi.15} and superfluid fraction~\cite{Zhe.16} measurements have been made in these alumina aerogels, the latter with a version called `obninsk', with evidence for a transition to the polar state of \SHE\ on cooling from the normal state.  The NMR frequency shift in nafen-243 was found to be significantly larger than the $A$ phase which led Dmitriev {\it et al.}~\cite{Dmi.15} to identify this superfluid as the polar state based on Eq.~\ref{ratios}. These results are consistent with the theory of Aoyama and Ikeda that the polar state should be  stable immediately  below $T_c$ followed  by the $A$ phase, or perhaps a polar distorted $A$ phase, in uniaxially anisotropic aerogels.~\cite{Aoy.06}   

The observation of a polar state is remarkable for several reasons.  Most importantly, this one-dimensional state does not occur in pure \HE.  Rather its  existence depends on anisotropy imposed by the aerogel framework.  Secondly, the polar state can support half-quantum vortices (HQV) predicted by Volovik and Mineev~\cite{Vol.76} and recently observed by Autti {\it et al.} in nafen-243 aerogel.~\cite{Aut.16}  More recently, they found very long lived quantum coherence in the magnon Bose Einstein condensate in this  aerogel for which the stability condition $d\Delta\omega/d\mathrm{cos}\beta<0$ is satisfied.~\cite{aut.17}  Using a less dense nafen-90 aerogel Dmitriev {\it et al.}~\cite{Dmi.15} found that  the polar phase transforms from a polar phase to a polar distorted $A$ phase at lower temperatures where preliminary results indicate~\cite{Aut.16} that the HQV survives.  These vortices should harbor Majorana zero modes similar to that predicted for pure  \SHE\ $A$ and that have been the subject of broad interest in condensed matter physics. 

\subsection*{Summary}

Unconventional superconductors have been a major focus in condensed matter physics including investigation of the symmetry of order parameters and their topology in a wide range of systems.  Superfluid \HE\ has a special role to play as a part of this effort where new superfluid phases  have been discovered  using engineered anisotropic materials to study complex symmetry breaking.  This work is the confluence of research on quantum liquids and materials physics on the nanoscale, that emphasizes the significant effects of correlated impurities on the quantum state of a fermion superfluid:  a realization of the paradigm, ``order in disorder''.~\cite{Min.12}

\subsection*{Acknowledgments}

The author acknowledges contributions from Jeevak Parpia and Jim Sauls with whom he shares the 2017 Fritz London Memorial Prize for work in this review; and contributions from Donald Sprague, Tom Haard, Jan Kycia, Yoonseok Lee, Guillaume Gervais, John Davis, Hyoungsoon Choi, Johannes Pollanen, Jia Li, Andrew Zimmerman, Man Nguyen, and Ben Moy; and discussions with Erkki Thuneberg, Grigory Volovik, Vladimir Mineev, Igor Fomin, Vladimir Dmitriev, Takao Mizusaki, Yutaka Sasaki, John Saunders, and Josh Wiman. Research was supported by the National Science Foundation, Division of Materials Research: DMR-1602542.

\end{document}